\documentclass[twocolumn]{revtex4}
\usepackage{graphicx}
\usepackage{fancyhdr}
\usepackage{amsmath,amsfonts,amssymb}

\DeclareGraphicsExtensions{.jpg, .JPG , .png , .pdf, .gif, .svg}

\usepackage{svg}

\usepackage{color}
\definecolor{red}{rgb}{1,0,0}
\definecolor{gre}{rgb}{0,1,0}
\definecolor{blu}{rgb}{0,0,1}

\newcommand{\sk}{\vskip0.2truecm}

\usepackage{hyperref}   % hyperlink dans le document 
\hypersetup{
  bookmarks=true,           % show bookmarks bar?
  unicode=false,          % non-Latin characters in Acrobat's bookmarks
  pdftoolbar=true,        % show Acrobat's toolbar?
  pdfmenubar=true,        % show Acrobat's menu?%
  pdffitwindow=false,     % window fit to page when opened
  pdfstartview={FitH},    % fits the width of the page to the window
  pdftitle={My title},    % title
  pdfauthor={Author},     % author
  pdfsubject={Subject},   % subject of the document
  pdfcreator={Creator},   % creator of the document
  pdfproducer={Producer}, % producer of the document
  pdfkeywords={keyword1} {key2} {key3}, % list of keywords
  pdfnewwindow=true,      % links in new window
  breaklinks=true,       %permet le retour �� la ligne dans les liens trop longs
  urlcolor= blue,        %couleur des hyperliens
  linkcolor= red,        %couleur des liens internes
  colorlinks=true,       % false: boxed links; true: colored links
  linkcolor=blue, %red,          % color of internal links
  citecolor=blue,        % color of links to bibliography
  filecolor=blue,      % color of file links
  urlcolor=blue           % color of external links
}

\begin{document}

\title{Langton's Ant from the Ant's referential}

%\date{\today}

%\date{}
%\author{}

%\address{}

\author{Thomas Cailleteau}

\affiliation{${}$ \\ Sant Job Skolaj-Lise,  42 Kerguestenen Straed, 56100 An Oriant, Breizh}
%%Laboratoire de Physique Subatomique et de Cosmologie, UJF, CNRS/IN2P3, INPG \\
%53, av. des Martyrs, 38026 Grenoble cedex, France}

\email{cailleteau@lpsc.in2p3.fr}

\begin{abstract}
The article can be considered as the first part of this work. From a cosmological point of view, the evolution of the Langton's ant on a 2D lattice is studied from the \textit{ant's framework} : the lattice is not considered as its whole but is \textit{build} as the ant moves and discovers a new case. The aim of this article is twofold. Firstly, to see if one can explain the emergent behaviour of the ant as an analogous sytem of a particle crossing an horizon. It is therefore another way to look if a discrete system may have some caracteristics of a black hole. Secondly, to describe the problem from the ant framework which shows some potential explanations of the problem. Pointing toward some directions for an explanation, the evolution of the density of one color is studied and commented.
% The aim of this article is therefore to show another path in the attempt of an explanation of the ant motion.

\end{abstract}

\maketitle

%Rho0 et thighway graph
%Revoir question attractor oscillations
%Attractor oui different value rho0 mais si pb petit ?

\section{Introduction}

Langton's ant is a two-dimensional universal Turing machine with a very simple set of rules but complex emergent behaviour. It was invented by Chris Langton \cite{Langton} in 1986 and runs on a square lattice of black and white cells. \sk

In the following, we will explain briefly the Langton's ant problem based on \cite{LangtonWiki}, then show and comment the numerical results regarding the densities, before ending by bringing some assumptions, paths in order to explain the evolution of the ant.

\subsection{Rules of the motion and modes of behavior}

Squares on a plane are colored variously either black or white. We arbitrarily identify one square as the ant. The ant can travel in any of the four cardinal directions at each step it takes. The ant moves according to the rules below:

\begin{itemize}
\item[$\bullet$] At a white square, turn $90^\circ$ right, flip the color of the square, move forward one unit. 
\item[$\bullet$] At a black square, turn $90^\circ$ left, flip the color of the square, move forward one unit
\end{itemize}

These simple rules lead to a complex behavior. Three distinct modes of behavior are apparent, when starting on a completely white grid.
\begin{enumerate}
\item Simplicity. During the first few hundred moves it creates very simple patterns which are often symmetric.
\item Chaos. After a few hundred moves, a big, irregular pattern of black and white squares appear. The ant traces a pseudo-random path until around 10,000 steps.
\item Emergent order. Finally the ant starts building a recurrent \textbf{highway} pattern of 104 steps that repeats indefinitely.
\end{enumerate}

All finite initial configurations tested eventually converge to the same repetitive pattern, suggesting that the highway is an attractor of Langton's ant, but no one has been able to prove that this is true for all such initial configurations. It is only known that the ant's trajectory is always unbounded regardless of the initial configuration (Cohen Kong theorem).

\begin{figure}[htb] 
\includegraphics[width=0.4\textwidth]{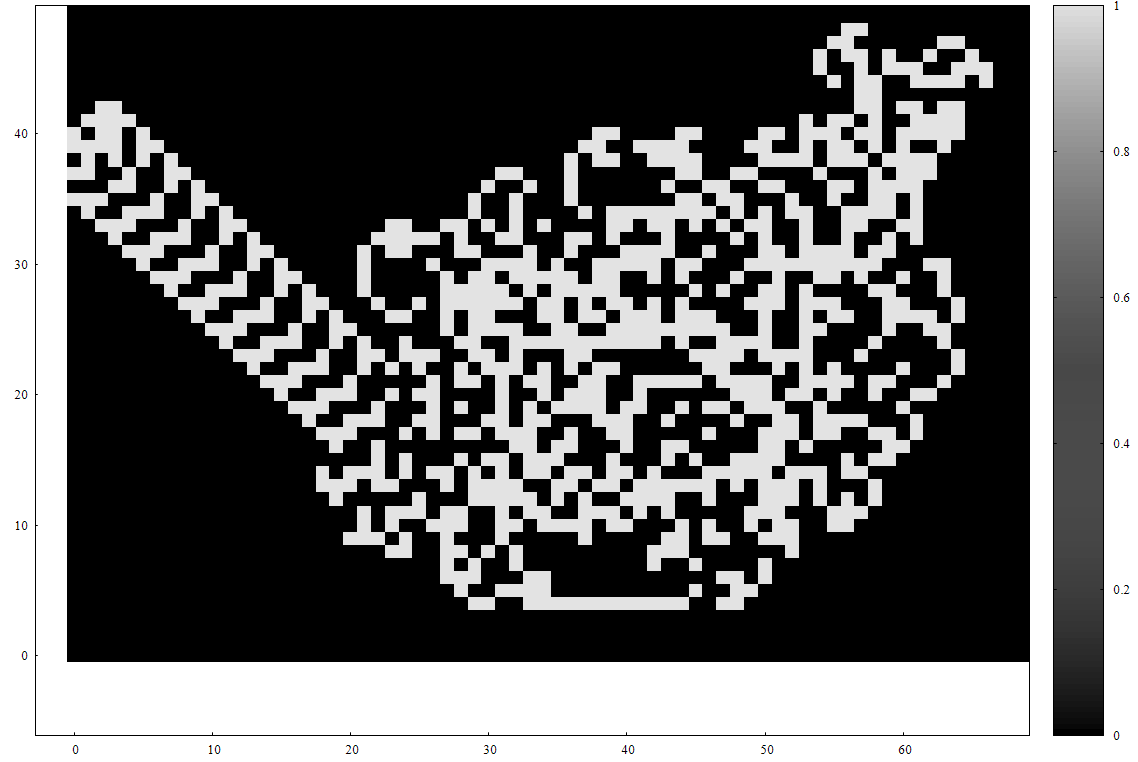}
\caption{Pattern of the ant's evolution observed after 12 000 steps (the colors are inverted)}
\label{fig:case1_lattice}
\end{figure}

\subsection{Definitions} \label{sec:definitions}

In the following, the terms \textit{step} and \textit{time} are interchangeable, and \textit{n-th step} and \textit{time n} have the same meaning.  In this work, the initial lattice was fully white (but the initial color does not matter).
\sk
As in Cosmology, the description of the evolution of an expanding/decreasing universe (as the bouncing scenario in Loop Quantum Gravity \cite{Ashtekar:2015dja}) is meaningful when studying the densities (of energy, of matter, for instance). In the case of the Langton's ant, it seems to be appropriate to proceed as such, as the particular evolution of the ant is given by the ant itself and the cases it has reached, and not by the whole universe at the end.
\sk

 Therefore, during this work, we have looked at
\begin{itemize}
\item $N(t)$, the total number of cases occupied at least once by the ant at a given time t (+1 when a new case is reached, $N(0)=1$). 
\item $B(t)$, the total number of black cases at some time/step t \textit{in the universe seen by the ant} \\ (+1/-1 when the ant changes the color of the case in black/white). When the case of a initally randomly filled lattice is considered, $B(0)=0$ if the initial case is white, 1 otherwise. 
\end{itemize}

The density of black cases is thus defined as
\begin{equation}
\rho(t) = \frac{B(t)}{N(t)}
\end{equation}

\section{Numerical observations : starting from a fully white lattice}

\subsection{Evolution of the density}

\begin{figure}[htb] 
\includegraphics[width=0.5\textwidth]{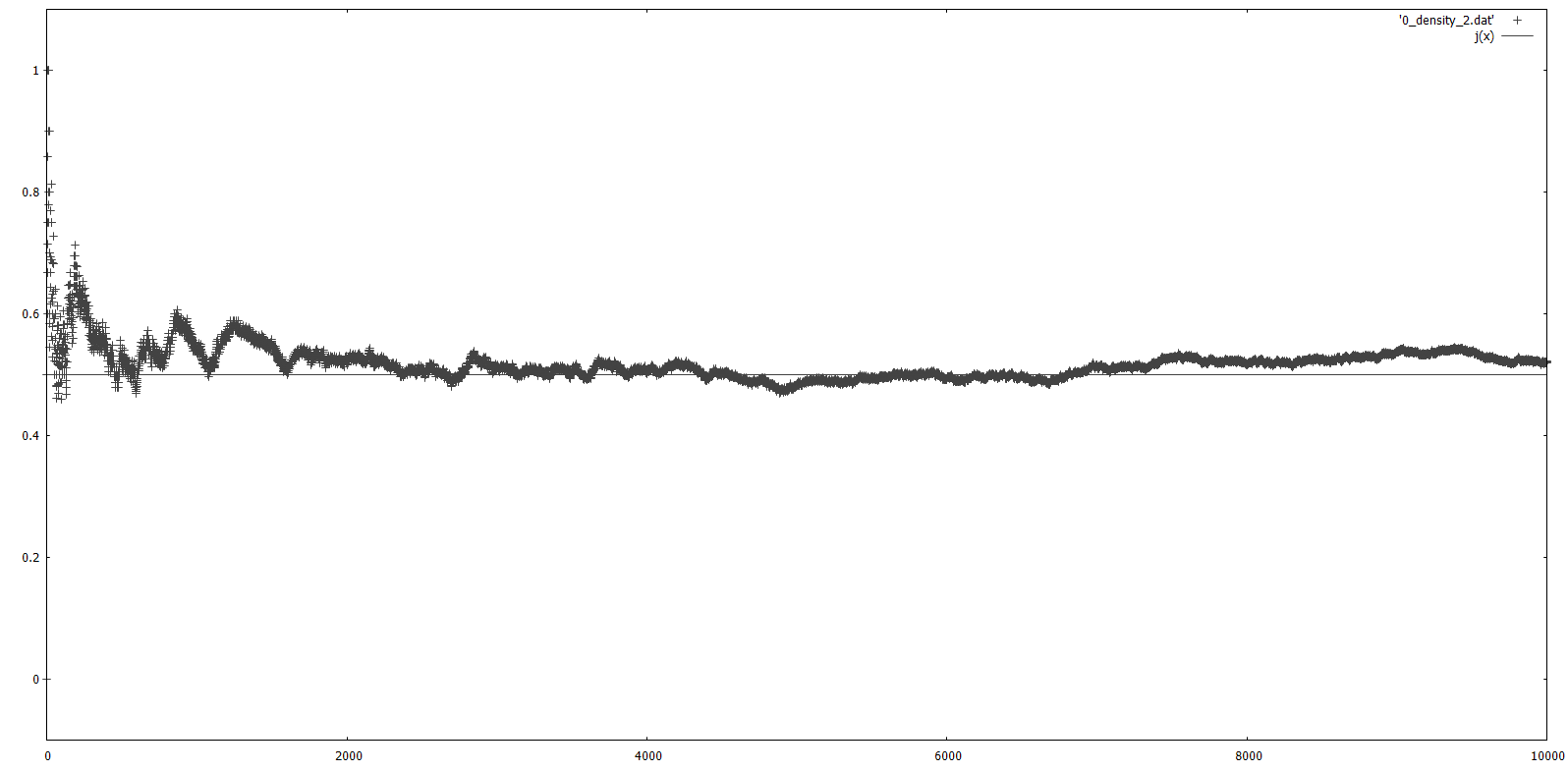}
\caption{Evolution of the density $\rho(t)$ as defined in Sec.(\ref{sec:definitions}) with the constant function f$(t) = 0.5$ : from t=1 to t=10 000. }
\label{fig:case1_energydens1}
\end{figure}

In Fig.(\ref{fig:case1_energydens1}), the density $\rho$ is shown until the transition area where the highway seems to start. If one zooms on the curve, a lot of small oscillations will be seen. 

\begin{figure}[htb] 
\includegraphics[width=0.5\textwidth]{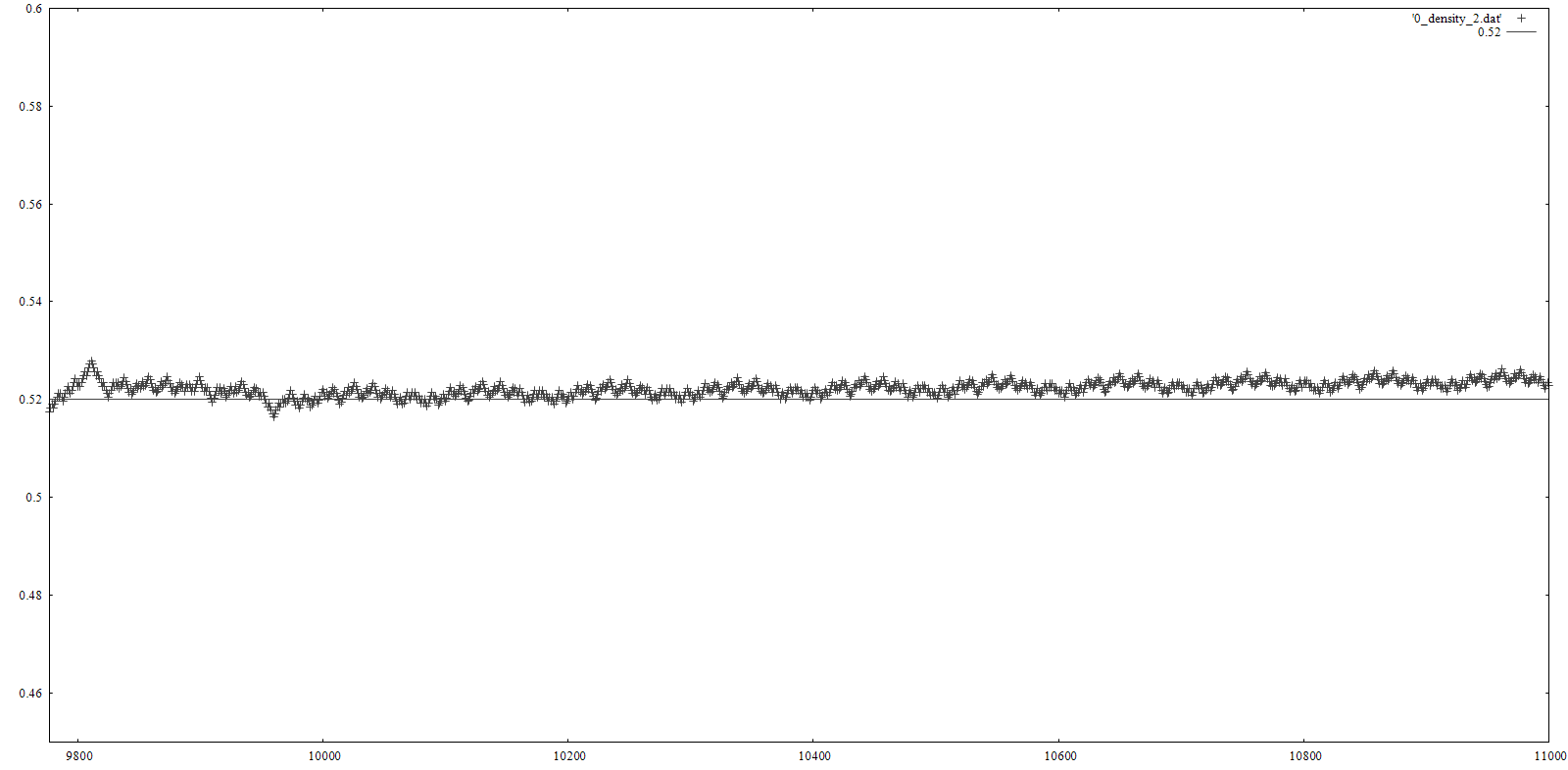}
\caption{Evolution of the density as defined in Sec.(\ref{sec:definitions}), during the Highway : : from t=9800 to t=11 000.}
\label{fig:case1_energydens1_part3}
\end{figure}

The evolution of the density during the highway is then represented in Fig.(\ref{fig:case1_energydens1_part3})

\subsection{Some comments}

\paragraph{First part : the stabilization} 
${}$ \sk

As seen in Fig.(\ref{fig:case1_energydens1}), neglecting the small oscillations and keeping only the general aspect of the curve, starting from a state where all case are white (state of low entropy as seen below), the ant seems not to be in a stable position, trying by its motion to equilibrate the amount of black and white tiles and therefore oscillating, analogously to a damped harmonic oscillator, around a value which is not exactly 0.5 but rather 0.52 here (attempts where the initial density $\rho_0 = \rho(0)$ was not 0 \textit{in an area around the initial position} lead to similar results as seen below).  Around t = 2200, the ant seems to enter an area where it has started to be much more stabilized, the \textit{force} which tries to maintain the density around 0.5 is more dominant and only a small deviation is observed (less than 10\%) until the transition area where the highway appears (t = 9990 approximatively as in Fig.(\ref{fig:case1_energydens1_part3})).

\vskip0.3truecm

\begin{figure}[htb] 
\includegraphics[width=0.5\textwidth]{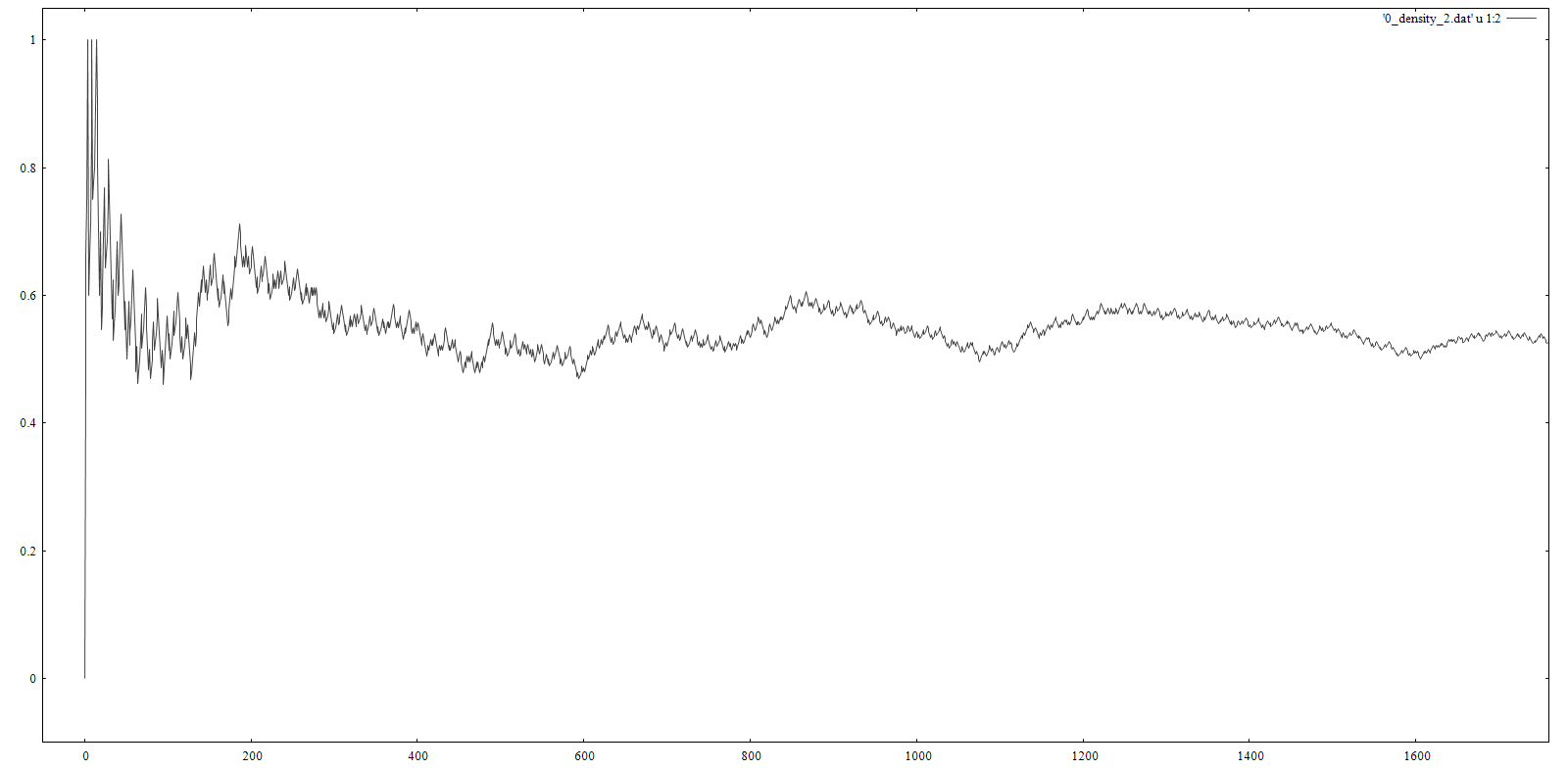}
\caption{Evolution of the density as defined in Sec.(\ref{sec:definitions}) until t = 1760. Zoom on the oscillations.}
\label{fig:case1_energydens1_part1}
\end{figure}

\begin{figure}[htb]
\includegraphics[width=0.5\textwidth]{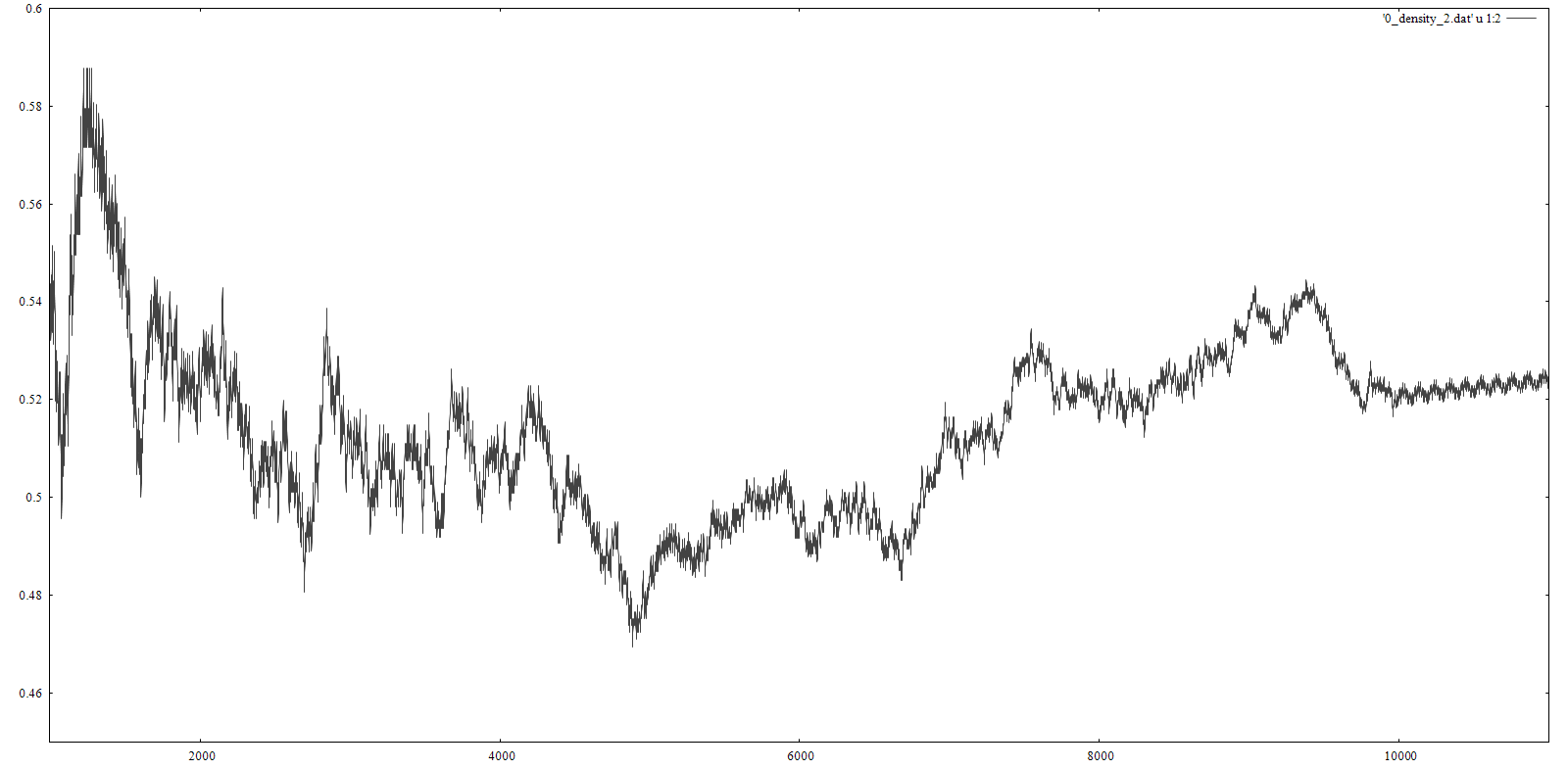}
\caption{Evolution of the density as defined in Sec.(\ref{sec:definitions}). Between t=1760 and 10000}
 \label{fig:case1_energydens1_part2}
\end{figure}

\paragraph{Second part : the Highway}
${}$ \sk

\begin{figure}[htb]
\includegraphics[width=0.5\textwidth]{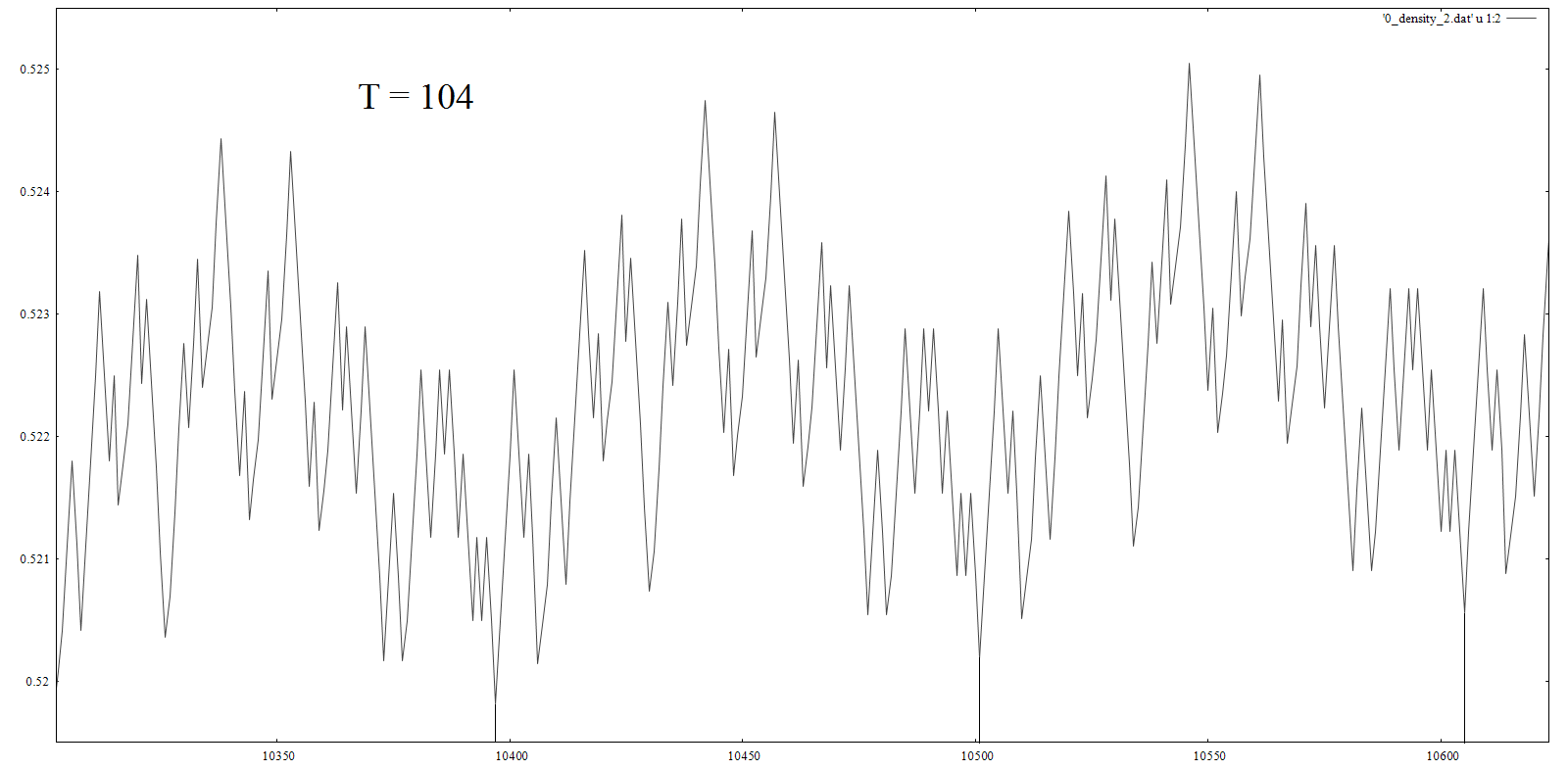}
\caption{Pattern of the density as defined in Sec.(\ref{sec:definitions}) during the Highway}
 \label{fig:case1_energydens1_pattern}
\end{figure}

As shown in Fig.(\ref{fig:case1_lattice}), during the highway, a pattern on the lattice is repeated indefinitely, leading to a certain pattern for the density, as seen in Fig.(\ref{fig:case1_energydens1_pattern}) : it should be \textit{easily} explained by studying the motion of the ant during the highway. Its period is also easily obtained : $T = 104$, which is coherent with the observed pattern of the highway.

\begin{figure}[htb] 
\includegraphics[width=0.5\textwidth]{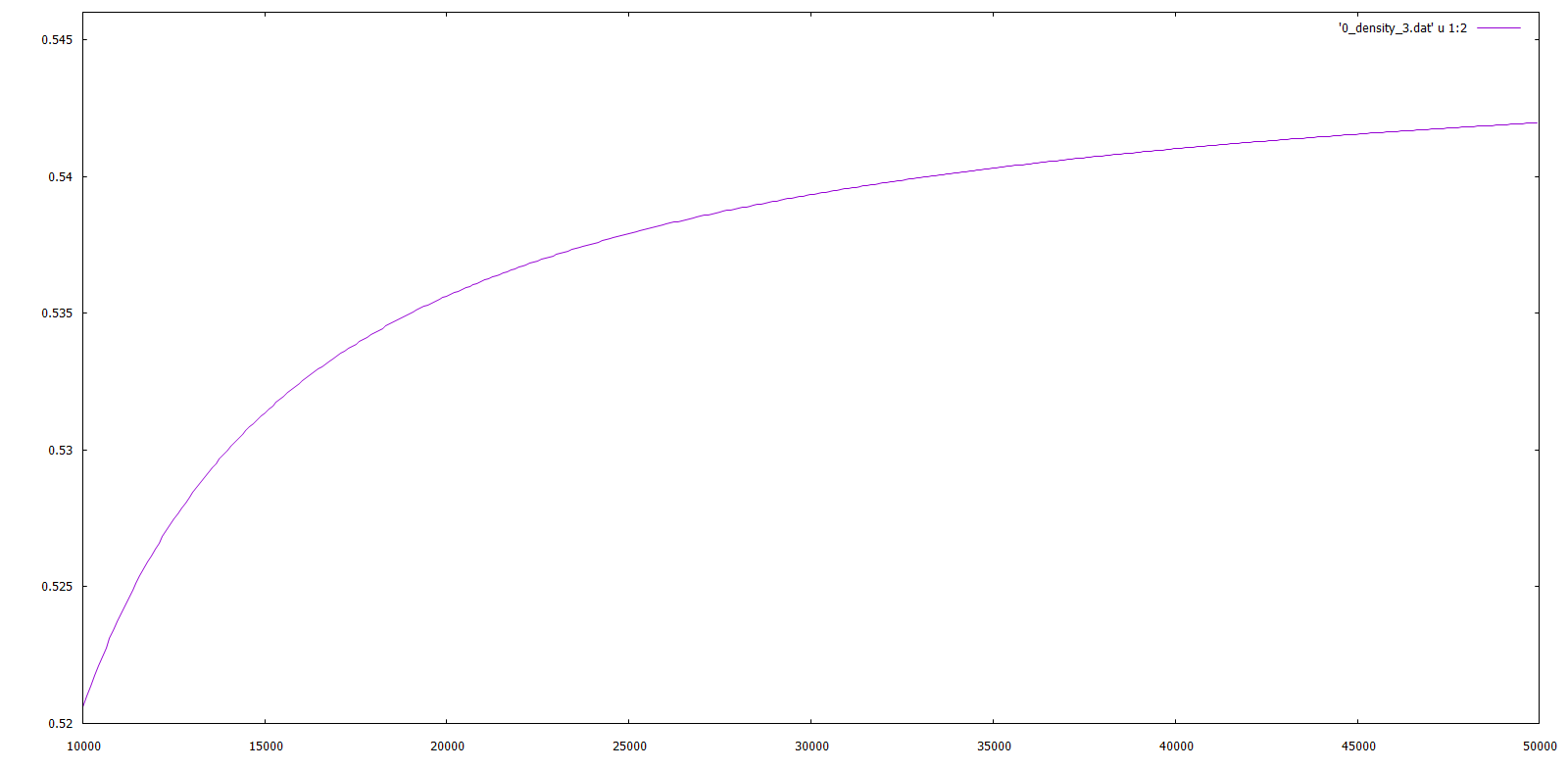}
\caption{Mean evolution of the density as defined in Sec.(\ref{sec:definitions}) during the Highway}
\label{fig:case1_highway_meanevolution}
\end{figure}

One may expect that the ant will have a behaviour such that the equilibrium $\rho = 0.5$ will be reached in some future. This is not actually the case, as  shown in the different figures. Moreover, after having taken the average values of $\rho(t)$ during the highway, one can see in Fig.(\ref{fig:case1_highway_meanevolution}) that the density increases at a really small rate : some \textit{good} linear regressions give for the highway : 

\begin{itemize}
\item for $t = 10 000$ to $t = 11 300$, \\ $\rho(t) = 3.184 \times 10^{-6} t + 0.48841$
\item for $t = 45 000$ to $t = 50 000$, \\  $\rho(t) = 8.3956 \times 10^{-8} t + 0.537773$
\end{itemize}

\begin{figure}[htb]
\includegraphics[width=0.5\textwidth]{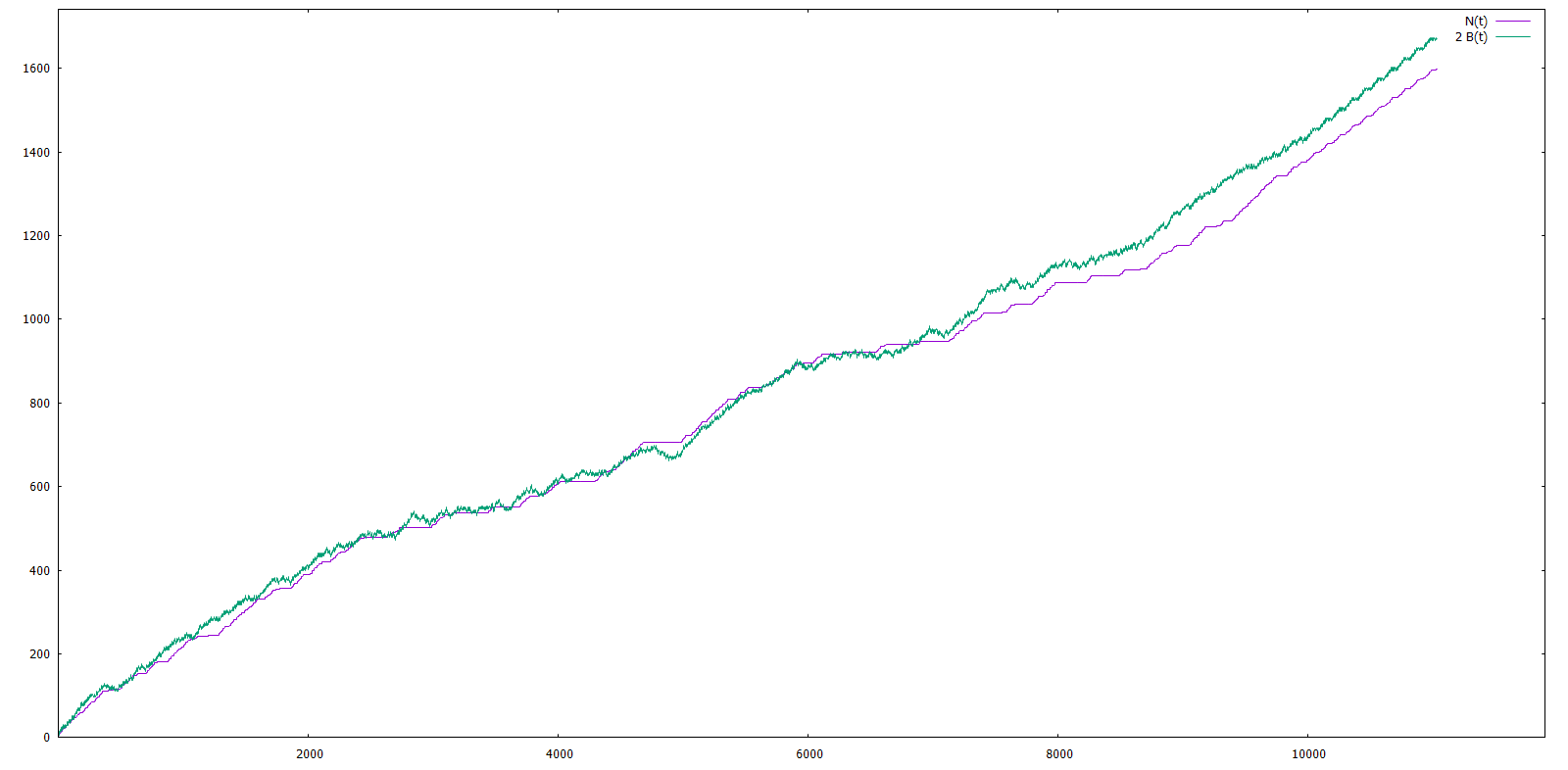}
\caption{Plot of $N(t)$ and $2 \times B(t)$. Between t=0 and 11900}
 \label{fig:case1_comp2BN}
\end{figure}

This is coherent as, during the highway, the ant creates more black cases than white ones, but also moves to more and more new tiles, as seen in Fig.(\ref{fig:case1_comp2BN}) where the evolution of $N(t)$ and $2 \times B(t)$ are plotted. Moreover, one can also see that until t=7500, both curves are \textit{relatively} closed, and starting from this time, both curves get away from each other. This leads to the question if this particular moment would initiate the emergent motion of the highway 2500 steps further.

. 

\paragraph{Some tests with a randomly filled lattice}
${}$ \sk

\begin{figure}[htb] 
\includegraphics[width=0.5\textwidth]{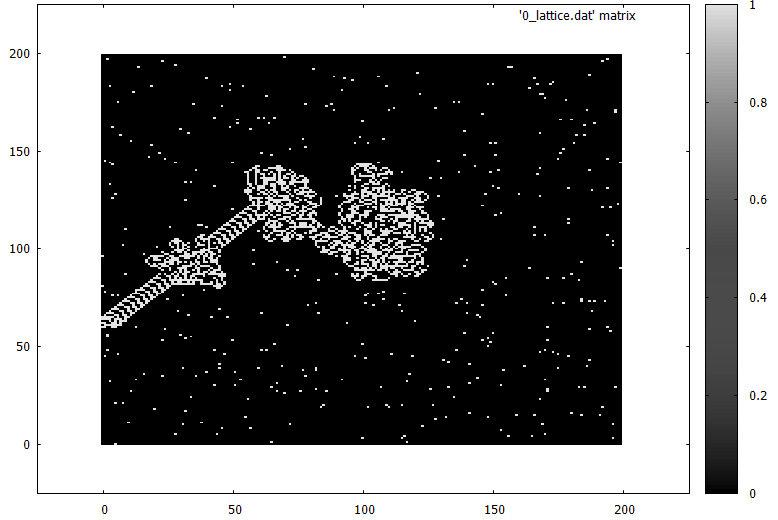}
\caption{Observed lattice when initially randomly filled with probabilities $p_B=0.01$ and $p_W=0.99$}
\label{fig:lattice2}
\end{figure}

\begin{figure}[htb]
\includegraphics[width=0.5\textwidth]{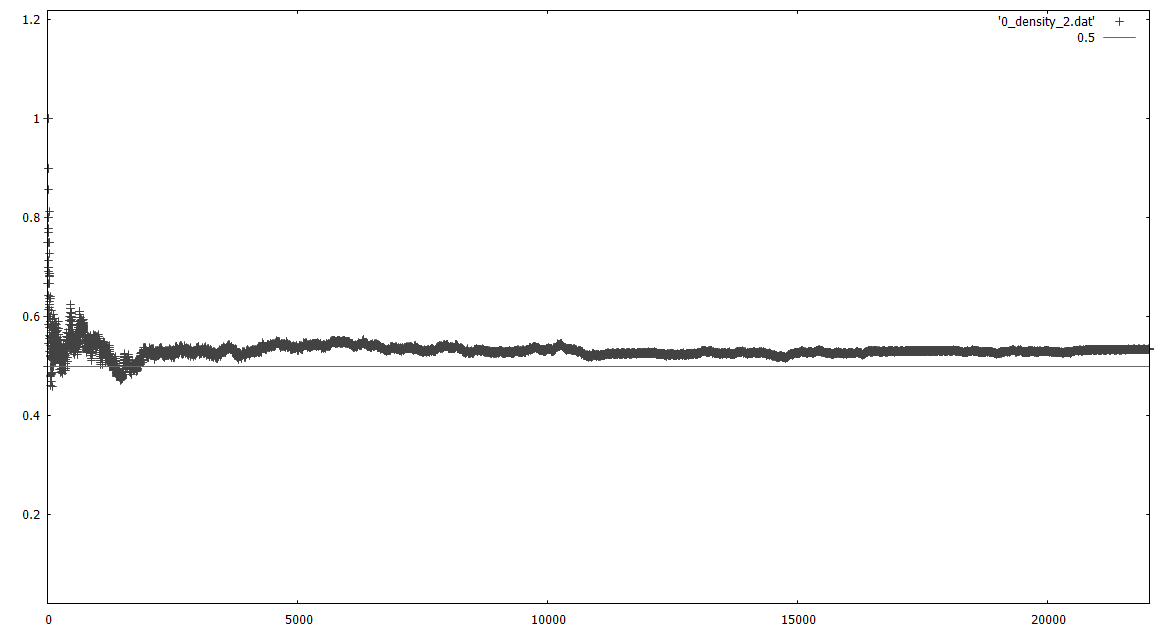}
\caption{Corresponding density $\rho(t)$ witht the lattice above, with the constant function 0.5. Probabilities $p_B=0.01$ and $p_W=0.99$.}
 \label{fig:case2_energydens1}
\end{figure}

We define $p_B$ and $p_W$ the probabilities that the case is initially black or white. Some tests have been made with randomly build lattices following the previous probabilities : the results as in Fig.(\ref{fig:lattice2}) and Fig.(\ref{fig:case2_energydens1}) seem to show similarities with the results obtained above, the density tends to reach a value around 0.5. It has been also observed that there seems to be a relation between the initial densities and the time when the highway appears. More work should be made in order to assert for sure such observations.

\section{The Langton's ant as a singularity}

\begin{figure}[htb]
\includegraphics[width=0.4\textwidth]{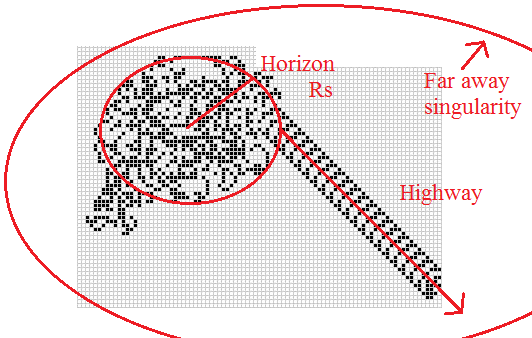}
\caption{Attempt to explain the emergent highway}
\label{fig:modelszh}
\end{figure}

The aim of this work was initially to see if one would be able to explain the emergent behavior of the ant with an analogous and physical model, namely as a point particle reaching a spacetime singularity. This is based on the idea that in our expanding universe, in the case where one considers it as spatially flat, after some calculations \cite{Carroll} knowing the amount of mass in the observed universe, one can see that the Schwarzschild radius $R_s$ corresponds to the Hubble radius, suggesting that we lives in a ... black hole. In the case of the Langton's ant, following such an idea, one can make the hypothesis that the emergent part, the highway, would appear when the density of  black (white) cases, combined with the radius of the area reached by the ant, would have reached a Schwarzschild-like value, leading the ant to cross the horizon and derive infinitely toward a singularity on the far edges of the universe as seen in Fig.(\ref{fig:modelszh}).

The Schwarzschild radius $R_S$ of a (spherical) object of mass $M$, is defined as 
\begin{equation}
R_S = \frac{2 \mathcal{G}}{c^2} M
\end{equation}
where $\mathcal{G}$ is the gravitational constant, and $c$ the velocity of light. If the case of the assumption where the mass M of the \textit{whole} system is constant in time, one has $\rho = \frac{M}{\mathcal{A}}$ where $\mathcal{A}$ is the area of the universe seen by the ant, then in this case, 
\begin{equation} \label{eq:scwz}
R_S \sim \rho_c \mathcal{A}_c
\end{equation}
where $\rho_c$ and $\mathcal{A}_c$ are the critical density and area when the highway seems to start.

\begin{figure}[htb] 
\includegraphics[width=0.4\textwidth]{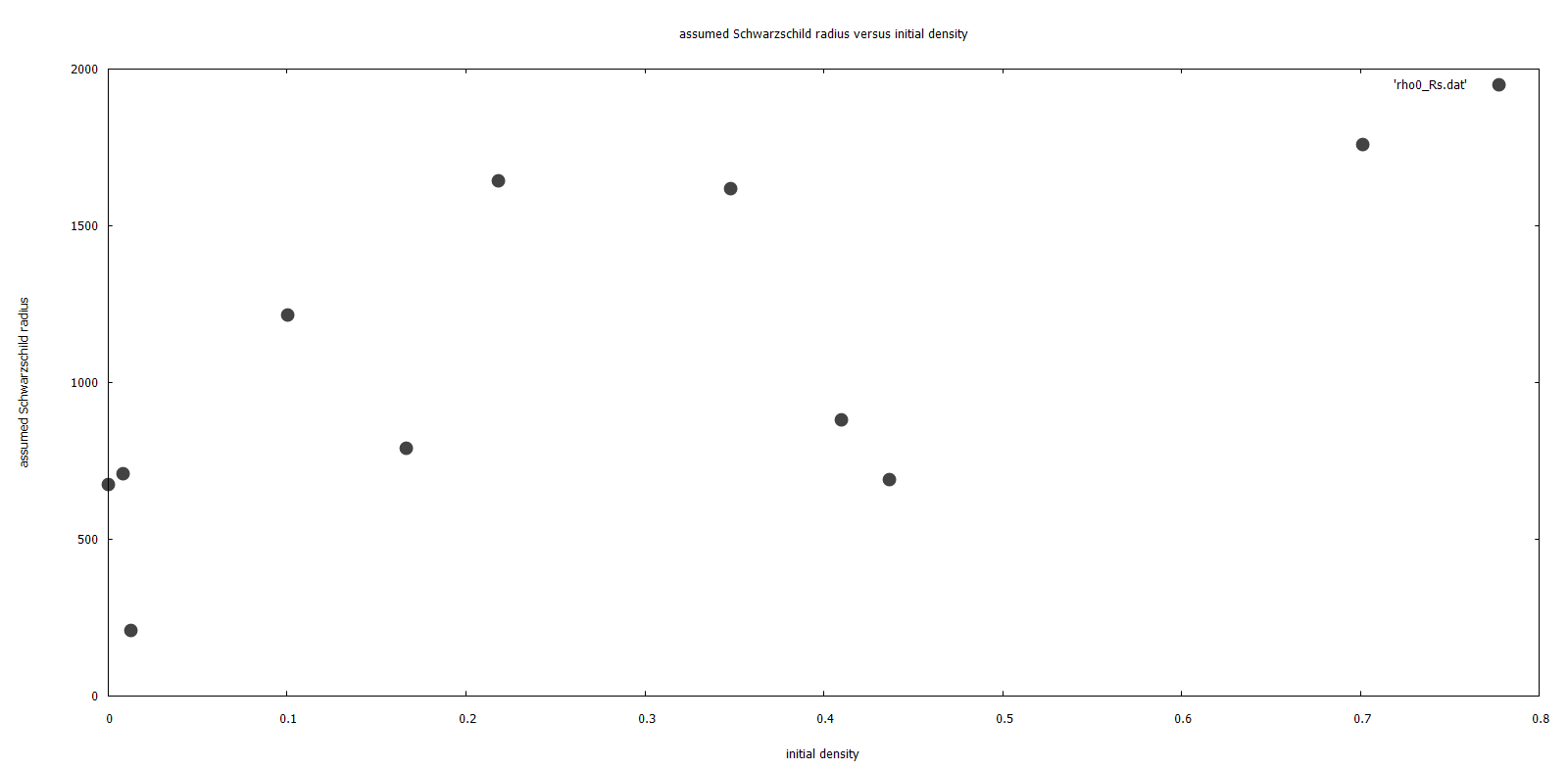}
\caption{Scwarzschild radius as in Eq.(\ref{eq:scwz}) versus initial density}
\label{fig:case1_RSrho0}
\end{figure}

\begin{figure}[htb] 
\includegraphics[width=0.4\textwidth]{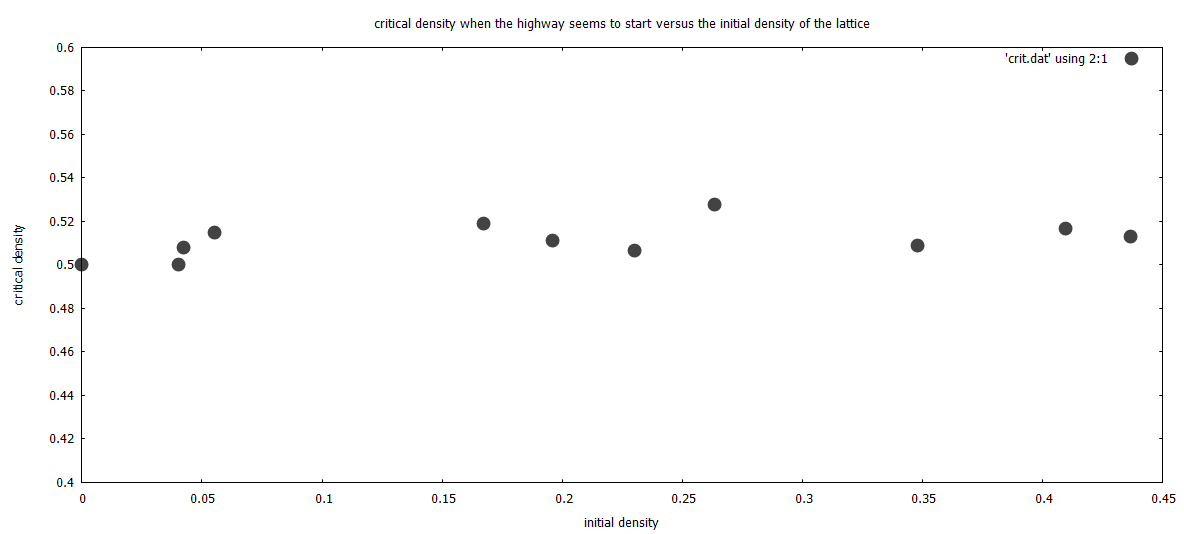}
\caption{critical density for the highway to start versus initial density}
\label{fig:case1_rhoc_rho0}
\end{figure}

 The comparison between $R_S$ as defined in Eq.(\ref{eq:scwz}) and the initial density $\rho_0$ is shown in Fig.(\ref{fig:case1_RSrho0}) where one can see that $R_S$ seems to increase as $\rho_0$ increases : this is then coherent with the fact that $\rho_0$ depends on $M$ and if $M$ increases then $\rho_0$ and $R_S$ increase too by the previous equations, and therefore $\rho_c$, as shown in Fig.(\ref{fig:case1_rhoc_rho0}). 
 
\begin{figure}[htb] 
\includegraphics[width=0.4\textwidth]{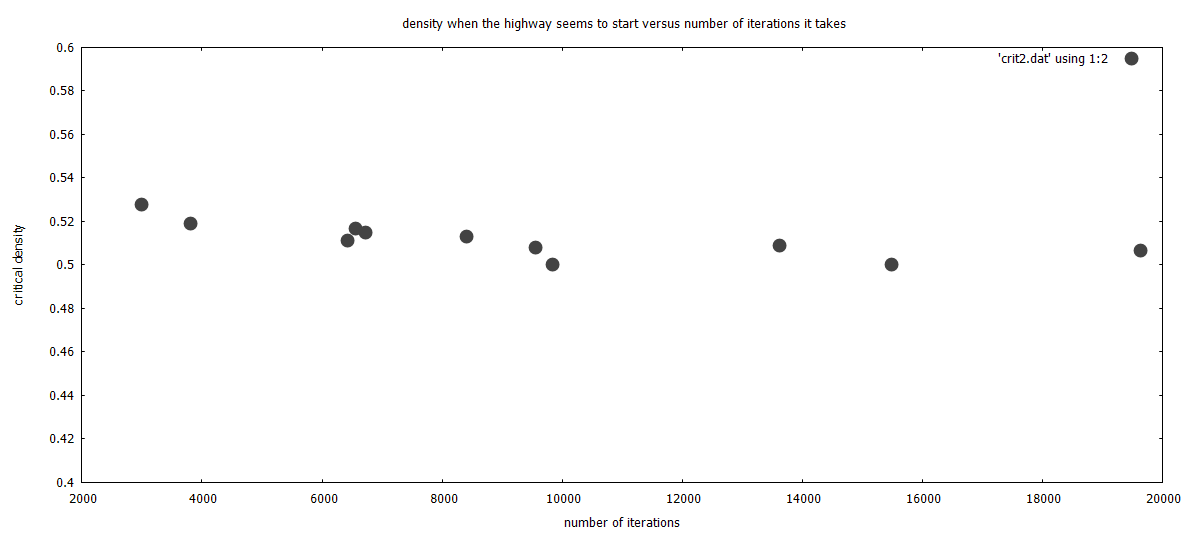}
\caption{critical density versus number of iterations it takes for the highway to start}
\label{fig:case1_rhocn}
\end{figure}

Moreover, as seen in Fig.(\ref{fig:case1_rhocn}), one can see that the longer it will take for the highway to appear, the less the critical density will be. 

The observed evolutions seem coherent. However, even if one can see some possible relations between the different observables, a statistical analysis is necessary to refine the model. Moreover, $M$ was define as the \textit{mass} of the system, and the definition of the mass for the Langton's ant is not yet defined but seems thus depend on the initial number of black cases. This work is a prospection of what could be done as a first approach, and more has to be achieved to see if it is really possible to understand the Langton's ant problem as a black hole \cite{LgtA2}.

\section{Some other hints / perspectives of explanations}

As in thermodynamics, equilibrium is generally obtained when the entropy of the system stays the highest. In the case of the Langton ant, states where the lattice is all black or white would be the ones with the lowest entropy, and consequently, states describing the situations where cases are roughly half black and half white would be the ones with the highest entropy. \sk 

Therefore, if for instance, we think about the lattice of Langton's ant as a system of particles with only two states available, black or white (or of half-spin), equilibrium would be obtained when densities $\rho(t)$ of both colors are (around) the expected value of 0.5 . 
\sk 

The curve for the density $\rho(t)$ as in Fig.(\ref{fig:case1_energydens1}), with the hypothesis that it is similar to a system whose initial state is at the lowest entropy (only white tiles) and trying to reach equilibrium, shows the caracteristics of a system whose density is attracted toward a value closed to 0.5 : a mathematical description of this evolution in the framework of the ant where this work has been done, would enforces, maybe proves and not only suggests, this behaviour of attractor for the highway, and therefore explains the whole evolution and leads to some new consequences.
\sk

Following the idea of this work, a similar approach as the one in \cite{Hamann} combined with the work done in \cite{Jiao140506} and with the help of \cite{LawlerLimic}, may bring some parts of an explanation for the emergent part of the motion. Indeed, it was shown in \cite{Jiao140506} that the expected value of $N(t)$, at the limit where the number of steps done is infinite, can be approximated by  

\begin{equation}
N(t) \rightarrow \frac{\pi t}{Log(t)} \hskip0.7truecm (t \rightarrow + \infty)
\end{equation} 

Consequently, the expected density $\rho$ would be impacted 

\begin{equation}
\rho(t) \sim \rho\left(\alpha(t) \frac{ t}{Log(t)}\right) \hskip0.7truecm (t \rightarrow + \infty)
\end{equation} 

\begin{figure}[htb] 
\includegraphics[width=0.5\textwidth]{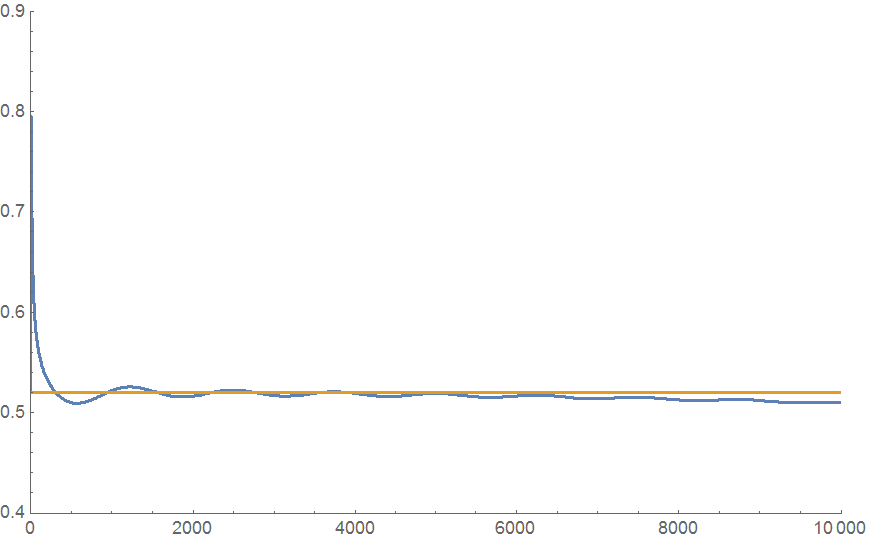}
\caption{Model given in Eq.(\ref{eq:model}) of the evolution of the density as defined in Sec.(\ref{sec:definitions})}
\label{fig:case1_energydens1_model}
\end{figure}

This expression is interesting : in a rough attempt, after having added some oscillatory features as in Eq.(\ref{eq:model}), it gives the curve in Fig.(\ref{fig:case1_energydens1_model}) which seems similar as the one in Fig.(\ref{fig:case1_energydens1_part2}) but need to be refined by a more rigorous approximation.

\begin{equation} \label{eq:model}
\rho(t) = \left(\frac{Log(t)}{t} Cos\left(\frac{t}{1000}\right) + 0.52 \right)  Cos\left(\frac{t}{50000}\right) 
\end{equation}

The results obtained in these works give only some approximations but the approaches may be suitable to give a description of the motion, at least a closed one which may help to understand better the phenomenon. 

\subsection{Some future possible works}

In order to refine the model, one may look at 
\begin{itemize}
\item some Fourrier analysis of the density : what is the corresponding power spectrum and what are the informations one would get ? 
\item some modifications of the rules, for instance, what does some asymetry in the model (example:  1 step on the right, 2 steps on the left) ? 
\item the caracteristics of the different evolutions of the ant in a lattice ramdomly filled with black and white tiles with the previous probabilities $p_B$ and $p_B$ : is it then possible to build a model which takes into account the values of $p_B$ and $p_W$ ? 
\item some approximations. For instance, the shape of the curve for the density is similar as the one for power spectrum of the tensorial perturbations in \cite{Mielczarek:2010bh}. 
\end{itemize}

\section{Conclusion}

Having a complete and rigorous explanation for the particular motion of the Langton's ant (and its other models) may be an objective quite hard to achieve. Nevertheless, this work shows the evolution of the ant from an other point of view, the one of a blind ant discovering the lattice step by step and only using few of it. 
\sk 

A first approach was then to look at the problem as a system analogous to a (Schwarzschild) black hole (another similar idea would have been to describe the ant as a model similar to a  star reaching the Chandrasekhar limit and collapsing to a supernovae). However, even if the idea seems interesting and this work shows some possibilities, it needs more inputs and work in order to explain it rigorously with such a model. Such a complete and consequent study is kept for the future.  
\sk
 Looking at the densities in this framework gives then some explanations from the point of view of entropy and equilibrium, but also, gives a way to explain the caracteristics of an attractor for the whole behaviour of the ant. In order to get such a suitable explanation, more analytical work is needed, and a suggestion would be to rely on some previous works and their results, being promising enough in this framework.
\sk

The Langton's ant problem and its variations are really interesting as cellular automatons, but it would also be interesting if such a mathematical problem would have an application in physics, as for instance in Quantum Gravity.

\begin{acknowledgments}
Python, Gnuplot and Mathematica were used here, the code has been modified from \url{https://rosettacode.org/wiki/Langton\%27s_ant}. The author would like to thanks Victor Santos for his help in Python, Florent Chevalier for having shown the Scratch version (bringing to the author a first idea of the Langton's ant as a singularity), Marc Suteau for giving the opportunity to work on this. 
\end{acknowledgments}

\end{document}